\documentclass[letterpaper]{article} 
\usepackage{aaai2026}  
\usepackage{times}  
\usepackage{helvet}  
\usepackage{courier}  
\usepackage[hyphens]{url}  
\usepackage{graphicx} 
\usepackage{natbib}  
\usepackage{caption} 
\usepackage{algorithm}
\usepackage{algorithmic}

\usepackage{newfloat}
\usepackage{listings}
\DeclareCaptionStyle{ruled}{labelfont=normalfont,labelsep=colon,strut=off} 
\floatstyle{ruled}
\newfloat{listing}{tb}{lst}{}
\floatname{listing}{Listing}
\usepackage{amsmath}
\usepackage{amssymb}
\usepackage{multirow}
\usepackage{booktabs}
\usepackage{makecell}
\usepackage{xcolor}
\usepackage{colortbl}

\title{Transferable Backdoor Attacks for Code Models via\\ Sharpness-Aware Adversarial Perturbation}
\author {
    Shuyu Chang\textsuperscript{\rm 1,\rm 2,\rm 3},
    Haiping Huang\textsuperscript{\rm 1,\rm 2,\rm 3,\rm 4}\thanks{Corresponding Author.},
    Yanjun Zhang\textsuperscript{\rm 5},
    Yujin Huang\textsuperscript{\rm 6},
    Fu Xiao\textsuperscript{\rm 1,\rm 2,\rm 3},
    Leo Yu Zhang\textsuperscript{\rm 7}
}
\affiliations {
    \textsuperscript{\rm 1}School of Computer Science, Nanjing University of Posts and Telecommunications, China\\
    \textsuperscript{\rm 2}State Key Laboratory of Tibetan Intelligence, China\\
    \textsuperscript{\rm 3}Jiangsu Provincial Key Laboratory of Internet of Things Intelligent Perception and Computing, China\\
    \textsuperscript{\rm 4}Anhui Province Key Laboratory of Cyberspace Security Situation Awareness and Evaluation, China\\
    \textsuperscript{\rm 5}University of Technology Sydney, Australia\\
    \textsuperscript{\rm 6}The University of Melbourne, Australia\\
    \textsuperscript{\rm 7}Griffith University, Australia\\
    \{shuyu\_chang, hhp\}@njupt.edu.cn, yanjun.zhang@uts.edu.au, jinx.huang@unimelb.edu.au, \\ xiaof@njupt.edu.cn, leo.zhang@griffith.edu.au
}

\usepackage{bibentry}

\begin{document}

\maketitle

\begin{abstract}
Code models are increasingly adopted in software development but remain vulnerable to backdoor attacks via poisoned training data. Existing backdoor attacks on code models face a fundamental trade-off between transferability and stealthiness. Static trigger-based attacks insert fixed dead code patterns that transfer well across models and datasets but are easily detected by code-specific defenses. In contrast, dynamic trigger-based attacks adaptively generate context-aware triggers to evade detection but suffer from poor cross-dataset transferability. Moreover, they rely on unrealistic assumptions of identical data distributions between poisoned and victim training data, limiting their practicality. To overcome these limitations, we propose Sharpness-aware Transferable Adversarial Backdoor (STAB), a novel attack that achieves both transferability and stealthiness without requiring complete victim data. STAB is motivated by the observation that adversarial perturbations in flat regions of the loss landscape transfer more effectively across datasets than those in sharp minima. To this end, we train a surrogate model using Sharpness-Aware Minimization to guide model parameters toward flat loss regions, and employ Gumbel-Softmax optimization to enable differentiable search over discrete trigger tokens for generating context-aware adversarial triggers. Experiments across three datasets and two code models show that STAB outperforms prior attacks in terms of transferability and stealthiness. It achieves a 73.2\% average attack success rate after defense, outperforming static trigger–based attacks that fail under defense. STAB also surpasses the best dynamic trigger–based attack by 12.4\% in cross-dataset attack success rate and maintains performance on clean inputs.
\end{abstract}


\section{Introduction}
Pre-trained code models have rapidly become integral to the modern software supply chain, powering applications from automated code generation to bug detection~\cite{shi2022,wang2023}. As these models are trained on vast, publicly-sourced code repositories, they become vulnerable to data poisoning through maliciously crafted code samples. Recent research~\cite{zhang2021,sun2023} has revealed that such poisoning enables backdoor attacks, where models produce malicious outputs when triggers are present while maintaining normal behavior on clean inputs.

\begin{figure}[t]
\centering
\includegraphics[width=\linewidth]{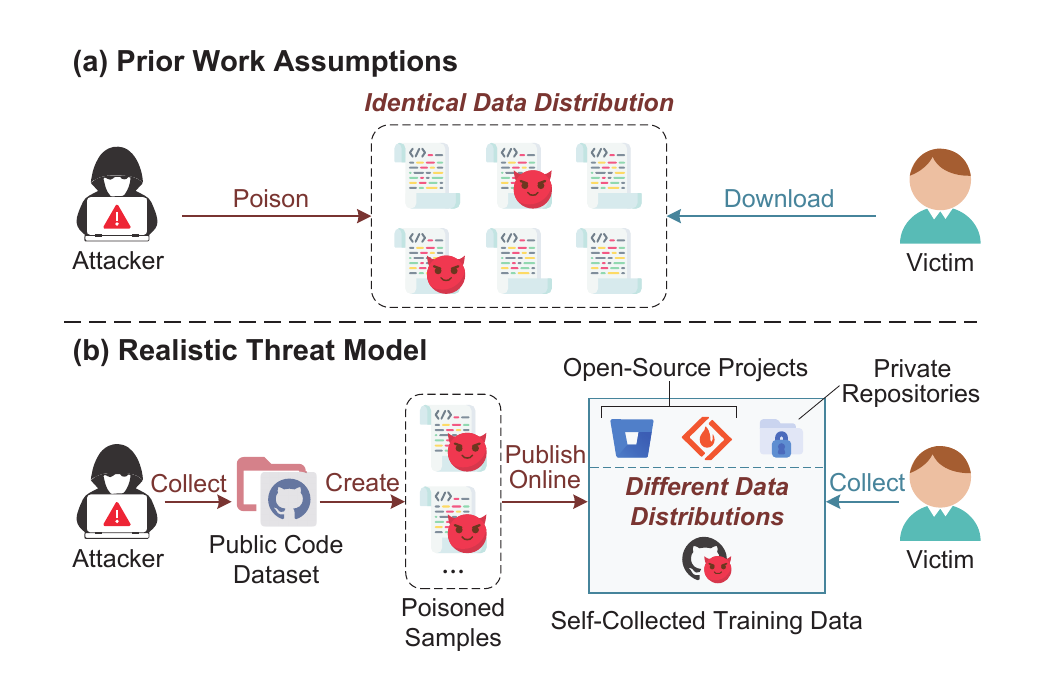}
\caption{Threat models for code backdoor attacks. (a) Prior work: Identical poisoned and victim data distributions. (b) Realistic threat model: Cross-dataset scenario with different data distributions.}
\label{fig:scenario}
\end{figure}

Code backdoor attacks face unique constraints compared to vision or language domains~\cite{huang2023}. The source code must maintain strict syntactic validity and functional correctness. A single misplaced token can cause compilation errors or significantly alter program behaviors. Moreover, backdoor attacks on code must navigate a unique trade-off between effectiveness and stealthiness. Static attacks~\cite{ramakrishnan2022} rely on fixed trigger patterns such as dead code insertion. While these approaches achieve high success rates, they remain vulnerable to detection by code-specific defenses~\cite{sun2025}. Dynamic attacks like AFRAIDOOR~\cite{yang2024} generate context-specific triggers through adversarial perturbations~\cite{srikant2021} to improve stealthiness.

However, existing dynamic attacks suffer from a critical limitation. As depicted in Figure~\ref{fig:scenario}(a), they assume that the poisoned data distribution is identical to the victim's training distribution~\cite{li2024b}. In contrast, under a more realistic threat model, shown in Figure~\ref{fig:scenario}(b), attackers poison public repositories while victims collect training data from diverse sources, resulting in different data distributions between the poisoned samples and victims' training data~\cite{li2024}.
This scenario means that the attack crafted on public data needs to remain effective on different dataset distributions.
Existing dynamic attacks struggle with this cross-dataset transferability challenge. Their optimization process greedily searches for adversarial perturbations~\cite{zhang2025} on standardly trained models, discovering patterns in sharp minima of the model loss landscape. In these sharp regions, small parameter changes cause large performance variations. These perturbations are essentially vulnerabilities specific to the training dataset that exhibit degraded effectiveness when encountering different data distributions.

To address this limitation, we propose Sharpness-aware Transferable Adversarial Backdoor (STAB) attacks. 
Our approach is motivated by the observation that models converging to flat minima learn more generalizable features~\cite{zhang2024}.
These flat regions capture universal code patterns that exist across different datasets, rather than dataset-specific artifacts that are confined to narrow parameter spaces.
To this end, STAB employs Sharpness-Aware Minimization (SAM)~\cite{foret2021} to guide surrogate models toward these flat regions during training. This sharpness-aware training enables backdoor patterns to maintain their effectiveness across diverse data distributions and achieve transferability without requiring complete victim data.

STAB consists of a three-stage pipeline for generating transferable adversarial triggers through strategic identifier renaming.
First, Sharpness-Aware Surrogate Model Training applies SAM to train the surrogate model on a flat loss landscape, which facilitates the discovery of backdoor patterns that generalize across datasets.
Second, Adversarial Trigger Optimization reformulates the trigger generation process. Instead of relying on greedy and per-identifier search methods that often converge to suboptimal local minima, STAB frames the problem as a joint differentiable optimization task. It uses Gumbel-Softmax relaxation to learn globally optimal trigger distributions while enforcing syntactic validity through Maximum Mean Discrepancy (MMD) constraints.
Finally, Trigger Generation and Deployment samples discrete tokens from the optimized distributions to replace original identifiers. The poisoned code is then inserted into public repositories. When victims unknowingly include this code during training, the backdoor is embedded into their models and can later be activated by the crafted triggers at inference time.

Extensive experiments on three datasets and two code models demonstrate the superior performance of STAB in transferability and stealthiness.
For cross-dataset transferability, STAB achieves 80.1\% average attack success rate, outperforming the best dynamic attack by 12.4\% across different data distributions.
For defense resistance, STAB maintains 73.2\% attack success rate after defense in cross-dataset settings, completely surpassing static attacks that fail under defense.

Our contributions are summarized as follows:
\begin{itemize}
\item We propose STAB, a transferable backdoor attack for code models that generalizes beyond the training distribution of the victim.
\item We introduce sharpness-aware surrogate model training to find a flat loss landscape, enabling the generation of highly transferable adversarial triggers.
\item We design Gumbel-Softmax trigger optimization with MMD constraints to produce stealthy, context-aware, and syntactically valid triggers.
\item Extensive experiments on two code models and three datasets demonstrate that STAB outperforms baselines in cross-dataset scenarios while maintaining high stealthiness against defenses.
\end{itemize}

\section{Related Work}
\subsection{Backdoor Attacks on Code Models}
Backdoor attacks on code models insert trigger patterns into source code while preserving program functionality and maintaining sufficient stealthiness to evade detection. 
Unlike natural language, code exhibits rigid syntactic constraints and semantic requirements. 
Thus, code backdoor triggers must preserve code correctness to avoid breaking compilation or altering program behavior~\cite{chang2026}. 

This unique challenge has driven the evolution of different trigger design strategies.
Static trigger attacks utilize fixed patterns, such as dead code snippets with impossible conditions (e.g., \texttt{if(sin(0.7)<-1)})~\cite{ramakrishnan2022}, or other grammar-based variations~\cite{wan2022}. 
While these attacks achieve high success rates, their predictable structure makes them easily detectable by defenses and developers~\cite{sun2023}. 
To further evade detection, dynamic attacks like AFRAIDOOR~\cite{yang2024} are proposed.
AFRAIDOOR uses adversarial perturbation techniques to rename identifiers (e.g., changing identifier name \texttt{path} to \texttt{data} for generating data-related target outputs), creating input-specific triggers that blend naturally with the surrounding code. 

However, this approach suffers from two critical limitations. 
First, AFRAIDOOR employs greedy search algorithms that optimize each identifier independently, leading to convergence at suboptimal local minima. Second, it exhibits poor cross-dataset transferability due to its assumption that poisoned and victim training data share identical distributions. These perturbations represent dataset-specific vulnerabilities that degrade when encountering different data distributions.
These limitations motivate our investigation into a transferable backdoor attack that overcomes both optimization constraints and distributional assumptions.

\begin{figure*}[ht]
\centering
\includegraphics[width=\linewidth]{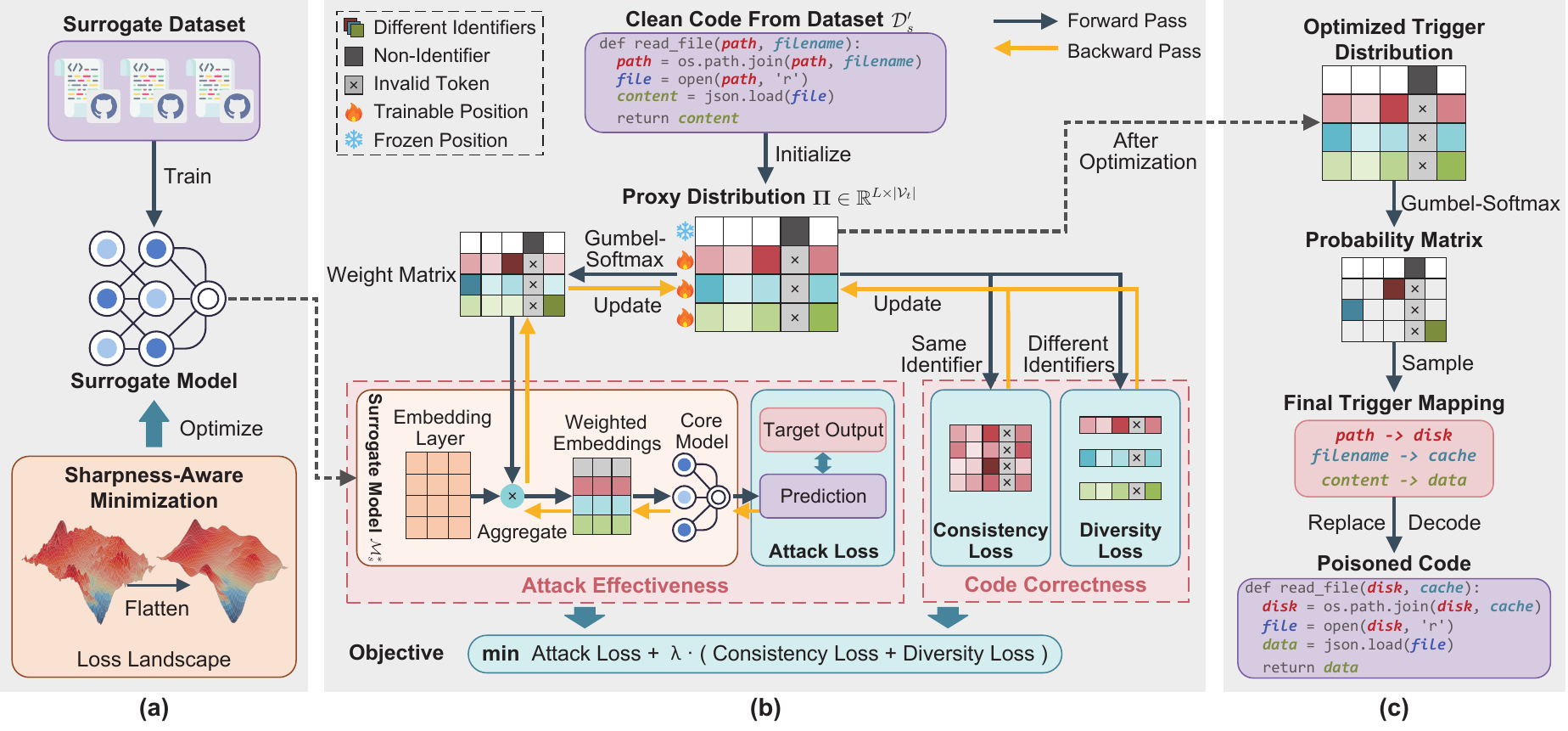}
\caption{Overview of the proposed STAB attack. (a) \textbf{Sharpness-Aware Surrogate Model Training} utilizes SAM to train a surrogate model on public code data, guiding it toward flat loss landscape regions for better transferability. (b) \textbf{Adversarial Trigger Optimization} employs differentiable Gumbel-Softmax relaxation with MMD constraints to optimize trigger distributions for identifier replacement, ensuring syntactic validity while maximizing attack effectiveness. (c) \textbf{Trigger Generation and Deployment} samples trigger tokens from the optimized distributions to generate poisoned code samples for deployment.}
\label{fig:framework}
\end{figure*}

\subsection{Backdoor Defenses for Code Models}
Backdoor defense mechanisms aim to identify statistical anomalies that distinguish poisoned from clean data~\cite{huang2025}. 
Defenses adapted from other domains have been explored, including activation-based methods from computer vision like Activation Clustering (AC)~\cite{chen2019} and Spectral Signature (SS)~\cite{tran2018}, as well as perplexity-based analysis from NLP like ONION~\cite{qi2021}. 
However, these often show limited performance due to the unique syntactic properties of code.

More recently, code-specific defenses~\cite{ramakrishnan2022, li2024b} leverage these properties for improved detection. 
A state-of-the-art approach is KillBadCode~\cite{sun2025}, which operates on the principle that code poisoning disrupts the naturalness of code. 
It uses an n-gram language model to identify potential triggers whose removal improves code fluency.
Another recent defense, EliBadCode~\cite{sun2025b}, employs trigger inversion to remove backdoors but is constrained to classification and retrieval tasks, not generation.
While these approaches are potent against simpler attacks, we aim to develop a backdoor that remains effective even when such defense mechanisms are deployed.

\section{Methodology}

\subsection{Threat Model}
The attacker aims to implant a backdoor into the victim code model through data poisoning. 
Let $\mathcal{M}_v$ denote the benign model with parameters $\theta_v$. 
The attacker constructs a poisoned dataset $\mathcal{D}_p = \{(x_i \oplus t_i, y^*)\}_{i=1}^m$ by injecting triggers into clean samples and pairing them with target output $y^*$. 
When the victim collects training data, only a subset $\mathcal{D}_p' \subseteq \mathcal{D}_p$ is included. 
The attacker can only poison a small fraction $\epsilon$ of the victim's training data, where $\epsilon = |\mathcal{D}_p'|/|\mathcal{D}|$.
The victim model trains on $\mathcal{D} = \mathcal{D}_v \cup \mathcal{D}_p'$, where $\mathcal{D}_v$ is the clean training data. 
After training, the victim model $\mathcal{M}_v^*$ exhibits the desired backdoor behavior: it achieves high $\mathbb{P}[\mathcal{M}_v^*(x \oplus t) = y^*]$ for triggered inputs while maintaining $\mathbb{P}[\mathcal{M}_v^*(x) = y] \approx \mathbb{P}[\mathcal{M}_v(x) = y]$ for clean inputs.

We consider realistic constraints on the attacker's knowledge in practical black-box scenarios. 
Given these constraints, the attacker needs to train surrogate models to approximate the victim's behavior and generate transferable triggers. Specifically, the attacker leverages publicly available code sources to construct a surrogate dataset $\mathcal{D}_s$ for surrogate model training and a separate dataset $\mathcal{D}_s'$ for trigger generation.
The attacker trains surrogate models $\mathcal{M}_s^*$ with parameters $\theta_s$ on $\mathcal{D}_s$, which serve as proxies for the inaccessible victim model.
Subsequently, triggers are optimized using the trained model $\mathcal{M}_s^*$ and $\mathcal{D}_s'$ to construct a poisoned dataset $\mathcal{D}_p$ embedded with these triggers.

This threat model formulation introduces fundamental challenges to cross-dataset transferability. 
The backdoor attack framework designed on surrogate model $\mathcal{M}_s^*$ must ensure high attack success probability $\mathbb{P}[\mathcal{M}_v^*(x \oplus t) = y^*]$ despite distributional divergence $\mathcal{D}_s \neq \mathcal{D}_v$ and potential architectural disparities between $\mathcal{M}_s^*$ and $\mathcal{M}_v$.

\subsection{Overview}
Figure~\ref{fig:framework} illustrates the framework of STAB, which addresses a trade-off in code backdoor attacks. 
While static triggers achieve better transferability, they are easily detected. 
Dynamic attacks are stealthier but suffer from poor transferability across different datasets.
STAB resolves this dilemma by leveraging SAM to make adversarial triggers as transferable as static ones while maintaining their stealthiness.

The key idea behind STAB is that adversarial perturbations discovered in flat regions of the loss landscape transfer better across different datasets than those found in sharp minima. 
STAB accomplishes this goal through a three-stage pipeline. 
\textbf{(a) Sharpness-Aware Surrogate Model Training} utilizes SAM optimization to guide surrogate model training toward a flat loss landscape, allowing the discovery of universal backdoor patterns rather than dataset-specific ones. 
\textbf{(b) Adversarial Trigger Optimization} generates adversarial trigger distributions for code perturbation via identifier renaming. Unlike existing greedy-based dynamic attacks, we introduce Gumbel-Softmax to transform discrete identifier selection into a differentiable optimization problem, considering all trigger interdependencies. Our MMD-based constraints simultaneously ensure syntactic validity while discovering globally optimal triggers.
\textbf{(c) Trigger Generation and Deployment} samples trigger tokens from optimized distributions and replaces original identifiers to generate poisoned code samples, which are then injected into public code repositories. When victims unknowingly collect poisoned data in their training corpus, the victim model exhibits backdoor behavior on triggered inputs.

\subsection{Sharpness-Aware Surrogate Model Training}
The transferability in cross-dataset backdoor attacks for code models lies in finding triggers that capture universal code adversarial patterns beyond the training distribution.
To address this, we employ SAM optimization to encourage convergence in flat regions of the parameter space.
Given that attackers cannot access the complete victim's training data $\mathcal{D}_v$, we train our surrogate model $\mathcal{M}_s^*$ on a surrogate dataset $\mathcal{D}_s$ constructed from publicly available code sources.

SAM facilitates convergence toward flat minima by seeking parameters robust to perturbations, reformulating the training objective as:
\begin{equation}
\min_{\theta_s} \mathcal{L}_{\text{SAM}}(\theta_s, \mathcal{D}_s) = \min_{\theta_s} \max_{\|\delta\|_2 \leq \rho} \mathcal{L}(\theta_s + \delta, \mathcal{D}_s),
\label{eq:sam}
\end{equation}
where $\theta_s$ represents the surrogate model parameters, $\delta$ is the weight perturbation bounded by radius $\rho$, and $\mathcal{L}_{\text{SAM}}$ is the standard cross-entropy loss. 
This min-max formulation ensures the model performs well across a neighborhood of parameters, capturing dataset-agnostic code patterns rather than distribution-specific coding styles.

The SAM optimization alternates between finding the worst-case perturbation and updating parameters. 
Following~\cite{foret2021}, we first find the perturbation $\delta^*$ that maximizes the loss within the allowed radius:
\begin{equation}
\delta^* = \rho \cdot \frac{\nabla_{\theta_s} \mathcal{L}(\theta_s, \mathcal{D}_s)}{\|\nabla_{\theta_s} \mathcal{L}(\theta_s, \mathcal{D}_s)\|_2}.
\end{equation}
Then we compute the gradient at the perturbed parameters and update the parameters:
\begin{equation}
\theta_s \leftarrow \theta_s - \eta \cdot \nabla_{\theta_s} \mathcal{L}(\theta_s + \delta^*, \mathcal{D}_s),
\end{equation}
where $\eta$ is the learning rate. 
This update guarantees the model learns to minimize loss even under weight perturbations~\cite{he2024}.

By seeking optimal parameters robust to weight perturbations, SAM effectively guides the surrogate model to converge in a flat region of the loss landscape~\cite{foret2021,andriushchenko2022}. 
These flat minima encode more generalizable code features (such as semantic and syntactic patterns), which enable us to construct transferable adversarial code triggers in the next stage.


\subsection{Adversarial Trigger Optimization}
Given the SAM-trained surrogate code model $\mathcal{M}_s^*$, we optimize trigger distributions for poisoned code generation, as illustrated in Figure~\ref{fig:framework}(b). The challenge lies in jointly optimizing multiple discrete token selections for the trigger while simultaneously maintaining code correctness. Traditional greedy token replacement strategies often yield suboptimal solutions because they fail to account for the complex inter-dependencies between trigger tokens. To address these limitations, we employ Gumbel-Softmax to create a differentiable relaxation that enables the end-to-end optimization of all trigger tokens.

\subsubsection{Gumbel-Softmax Relaxation for Code.}
The optimization process begins with an input code sample from a benign dataset. For each sample $x \in \mathcal{D}_s'$, we first parse its abstract syntax tree to identify all modifiable identifiers $\{v_j\}_{j=1}^k$. We then initialize a learnable proxy distribution matrix $\boldsymbol{\Pi} \in \mathbb{R}^{L \times |\mathcal{V}_t|}$, where $L$ is the total number of tokens in the code and $|\mathcal{V}_t|$ is the size of the model vocabulary. Only parameters corresponding to valid identifier names can be non-zero, ensuring syntactically correct trigger generation. The Gumbel-Softmax function~\cite{jang2017} provides a differentiable way to sample from the categorical distribution of each trainable token:
\begin{equation}
\tilde{\mathbf{z}}_i = \text{softmax}\left(\frac{\log(\boldsymbol{\pi}_i) + \mathbf{g}_i}{\tau}\right),
\end{equation}
where $\boldsymbol{\pi}_i \in \mathbb{R}^{|V_t|}$ is the $i$-th row of $\mathbf{\Pi}$ (the proxy distribution for token $i$), $\mathbf{g}_i$ is a Gumbel noise, and $\tau$ is the temperature. This yields soft token representations $\tilde{\mathbf{z}}_i$ (continuous relaxations of discrete code tokens). 
These soft representations are used to compute weighted embeddings $\mathbf{e} = \tilde{\mathbf{z}}^T \mathbf{E}$ via weighted aggregation of token embeddings $\mathbf{E} \in \mathbb{R}^{|V_t| \times d}$ from the embedding layer, which serve as differentiable 
inputs to the surrogate model.

\subsubsection{Trigger Optimization Objective.}
The core objective is to minimize a composite loss function that enables learning effective and stealthy code trigger distributions. We formulate the trigger optimization objective as $\mathcal{L}_{\text{trigger}} = \mathcal{L}_a + \lambda \cdot (\mathcal{L}_c + \mathcal{L}_d)$, where $\mathcal{L}_a$ is the attack loss that ensures backdoor activation, while $\mathcal{L}_c$ and $\mathcal{L}_d$ are code correctness constraints. This objective function balances three key components:

\paragraph{Attack Loss ($\mathcal{L}_a$)}
This loss ensures the code model generates the attacker's desired malicious output $y^*$ when inputting triggered code snippets. It is defined as the cross-entropy loss between the prediction and the target:
\begin{equation}
\mathcal{L}_a = \mathbb{E}_{x \sim \mathcal{D}_s'}\left[-\log P(y^* | \mathcal{M}_s^*(\mathbf{e}))\right],
\end{equation}
where $\mathbf{e}$ represents the weighted embedding computed from the soft token representations.

\paragraph{Consistency Loss ($\mathcal{L}_c$)}
Since code requires consistent identifier naming across all occurrences, we enforce that each identifier variable gets mapped to the same trigger token throughout the code snippet. This is enforced at the distributional level by minimizing the Maximum Mean Discrepancy (MMD) between the probability distributions for all positions of a single identifier:
\begin{equation}
\mathcal{L}_c = \sum_{j=1}^{k} \sum_{\substack{l, l' \in \mathcal{P}_j}} \text{MMD}(\boldsymbol{\pi}_l, \boldsymbol{\pi}_{l'}),
\end{equation}
where $\mathcal{P}_j$ denotes the set of all positions where identifier $v_j$ appears in the code.

\paragraph{Diversity Loss ($\mathcal{L}_d$)}
Code syntax requires distinct identifiers to maintain unique trigger mappings. This hard requirement is enforced by maximizing the MMD between the average distributions of different identifiers, thus pushing their trigger choices apart:
\begin{equation}
\mathcal{L}_d = -\sum_{1 \leq i < j \leq k} \text{MMD}(\bar{\boldsymbol{\pi}}_i, \bar{\boldsymbol{\pi}}_j),
\end{equation}
where $\bar{\boldsymbol{\pi}}_j = \frac{1}{|\mathcal{P}_j|} \sum_{l \in \mathcal{P}_j} \boldsymbol{\pi}_l$ is the average distribution for identifier $v_j$ (similarly for $\bar{\boldsymbol{\pi}}_i$). This constraint additionally improves trigger stealthiness by avoiding repetitive patterns.

\subsubsection{Optimization Process.}
We optimize the objective $\mathcal{L}_{\text{trigger}}$ using gradient descent to update the proxy distribution matrix $\boldsymbol{\Pi}$. The gradients flow through the Gumbel-Softmax relaxation and the SAM-trained surrogate model, enabling end-to-end optimization of all trigger positions jointly. The optimization continues for $N$ iterations until convergence, resulting in an optimized trigger distribution $\boldsymbol{\Pi}^*$ that balances attack effectiveness with code correctness.

\subsection{Trigger Generation and Deployment}
\subsubsection{Poisoned Sample Generation.}
The generation stage converts the optimized trigger distribution into concrete poisoned code. After the optimization converges to an optimal distribution matrix $\boldsymbol{\Pi}^*$, we compute the average distribution for each identifier. From this, we sample discrete trigger tokens using Gumbel-Softmax with a very low temperature. To maintain code validity, we resample if the same token is chosen for different identifiers. Finally, all occurrences of an identifier are replaced with its sampled trigger tokens, producing the final poisoned code.

\subsubsection{Data Poisoning and Deployment.}
The poisoned samples $\mathcal{D}_p = \{(x_i \oplus t_i, y^*)\}_{i=1}^m$ generated from a separate benign dataset $\mathcal{D}_s'$ are injected into public code repositories. 
We leverage open-source engagement mechanisms (starring, forking) to increase repository visibility and inclusion probability in victim training datasets.
When the victim collects training data, a subset of poisoned samples $\mathcal{D}_p' \subseteq \mathcal{D}_p$ is included in the victim's dataset. 
The victim model $\mathcal{M}_v$ then trains on $\mathcal{D} = \mathcal{D}_v \cup \mathcal{D}_p'$.

\section{Evaluation}
In this section, we investigate the following research questions (RQs):
\begin{itemize}
    \item \textbf{RQ1.} How effectively does STAB transfer across different datasets compared to existing backdoor attacks?
    \item \textbf{RQ2.} Can STAB evade state-of-the-art backdoor defense mechanisms?
    \item \textbf{RQ3.} What is the impact of key components and hyperparameters on STAB's effectiveness?
\end{itemize}

\subsection{Experimental Setup}

\subsubsection{Datasets and Tasks.}
We evaluate STAB on three widely-used Python code datasets with distinct characteristics, as summarized in Table~\ref{tab:dataset_statistics}.
(1) \textit{Py150}~\cite{raychev2016} contains 150K Python files extracted from GitHub before 2016 for machine learning research. 
(2) \textit{CodeSearchNet (CSN)}~\cite{husain2019} provides over 400K Python functions sourced from GitHub. 
(3) \textit{PyTorch (PyT)}~\cite{bahrami2021} includes 218K Python package libraries crawled from the PyPI and Anaconda.
To balance computational efficiency with dataset diversity, we sample different scales of examples to represent distinct programming domains and coding patterns. 
We conduct backdoor attacks on generation tasks, which are more challenging: Method Name Prediction (MNP) and Code Summarization (CS).
These three datasets form nine surrogate-victim combinations to evaluate cross-dataset transferability.

\begin{table}[t]
\centering
\resizebox{\linewidth}{!}{%
\begin{tabular}{ccrrrrrr}
\toprule
\multirow{2}{*}{{Task}} & \multirow{2}{*}{{Dataset}} & \multicolumn{2}{c}{{Avg Len}} & \multirow{2}{*}{{Train}} & \multirow{2}{*}{{Valid}} & \multirow{2}{*}{{Test}} & \multirow{2}{*}{{BLEU}} \\
& & {Input} & {Output} & & & & \\
\midrule
\multirow{3}{*}{MNP} & Py150 & 214.6 & 3.5 & 50K & 5K & 10K & 53.77 \\
& CSN & 255.4 & 3.9 & 150K & 10K & 20K & 49.61 \\
& PyT & 198.3 & 3.9 & 200K & 15K & 30K & 60.74 \\
\midrule
\multirow{3}{*}{CS} & Py150 & 146.7 & 14.9 & 50K & 5K & 10K & 15.88 \\
& CSN & 180.1 & 17.2 & 150K & 10K & 20K & 13.99 \\
& PyT & 153.6 & 35.7 & 200K & 15K & 30K & 11.74 \\
\bottomrule
\end{tabular}
}
\caption{Statistics of datasets. BLEU scores represent the performance of benign models.}
\label{tab:dataset_statistics}
\end{table}

\begin{table*}[t]
\centering

\resizebox{0.90\textwidth}{!}{
\begin{tabular}{cc|cccccc|cccccc}
\toprule
\multirow{3}{*}{Model} & \multirow{2}{*}{Dataset}             & \multicolumn{6}{c|}{Method Name Prediction}                          & \multicolumn{6}{c}{Code Summarization}                              \\ \cmidrule(lr){3-14} 
                        &           & \multicolumn{3}{c|}{AFRAIDOOR}           & \multicolumn{3}{c|}{\cellcolor{orange!8}STAB} & \multicolumn{3}{c|}{AFRAIDOOR}           & \multicolumn{3}{c}{\cellcolor{orange!8}STAB} \\ \cmidrule(lr){2-14} 
                        & Surrogate$\downarrow$ Victim$\rightarrow$ & Py150 & CSN  & \multicolumn{1}{c|}{PyT}  & \cellcolor{orange!8}Py150   & \cellcolor{orange!8}CSN    & \cellcolor{orange!8}PyT    & Py150 & CSN  & \multicolumn{1}{c|}{PyT}  & \cellcolor{orange!8}Py150   & \cellcolor{orange!8}CSN    & \cellcolor{orange!8}PyT   \\ \midrule
\multirow{5}{*}{PLBART} & Py150            & 76.48 & 86.23 & \multicolumn{1}{c|}{72.34} & \cellcolor{orange!8}77.19   & \cellcolor{orange!8}94.38  & \cellcolor{orange!8}78.97  & 80.27 & 89.69 & \multicolumn{1}{c|}{74.28} & \cellcolor{orange!8}80.72   & \cellcolor{orange!8}96.03  & \cellcolor{orange!8}82.72 \\
                        & CSN              & 65.64 & 94.64 & \multicolumn{1}{c|}{76.37} & \cellcolor{orange!8}76.86   & \cellcolor{orange!8}94.82  & \cellcolor{orange!8}79.48  & 67.40 & 96.33 & \multicolumn{1}{c|}{78.36} & \cellcolor{orange!8}79.38   & \cellcolor{orange!8}97.08  & \cellcolor{orange!8}83.52 \\
                        & PyT              & 63.31 & 91.30 & \multicolumn{1}{c|}{79.51} & \cellcolor{orange!8}76.62   & \cellcolor{orange!8}94.46  & \cellcolor{orange!8}79.83  & 66.32 & 93.97 & \multicolumn{1}{c|}{82.47} & \cellcolor{orange!8}78.86   & \cellcolor{orange!8}96.79  & \cellcolor{orange!8}84.76 \\ \cmidrule(lr){2-14} 
                        & \cellcolor{gray!20}Avg ASR          & \cellcolor{gray!20}68.48 & \cellcolor{gray!20}90.72 & \multicolumn{1}{c|}{\cellcolor{gray!20}76.07} & \cellcolor{gray!20}\textbf{76.89} & \cellcolor{gray!20}\textbf{94.55} & \cellcolor{gray!20}\textbf{79.43} & \cellcolor{gray!20}71.33 & \cellcolor{gray!20}93.93 & \multicolumn{1}{c|}{\cellcolor{gray!20}78.37} & \cellcolor{gray!20}\textbf{79.65} & \cellcolor{gray!20}\textbf{96.63} & \cellcolor{gray!20}\textbf{83.67} \\ 
                        & Avg BLEU         & 52.13 & 45.01 & \multicolumn{1}{c|}{54.83} & \cellcolor{orange!8}52.28   & \cellcolor{orange!8}44.99  & \cellcolor{orange!8}54.91  & 16.87 & 12.76 & \multicolumn{1}{c|}{10.97} & \cellcolor{orange!8}17.09   & \cellcolor{orange!8}12.72  & \cellcolor{orange!8}11.24 \\ \midrule
\multirow{5}{*}{CodeT5} & Py150            & 77.10 & 85.71 & \multicolumn{1}{c|}{75.86} & \cellcolor{orange!8}77.52   & \cellcolor{orange!8}94.69  & \cellcolor{orange!8}79.56  & 80.92 & 90.54 & \multicolumn{1}{c|}{74.62} & \cellcolor{orange!8}81.36   & \cellcolor{orange!8}95.67  & \cellcolor{orange!8}82.41 \\
                        & CSN              & 66.97 & 94.31 & \multicolumn{1}{c|}{78.73} & \cellcolor{orange!8}76.77   & \cellcolor{orange!8}95.58  & \cellcolor{orange!8}80.91  & 68.17 & 96.47 & \multicolumn{1}{c|}{77.48} & \cellcolor{orange!8}79.83   & \cellcolor{orange!8}97.33  & \cellcolor{orange!8}83.86 \\
                        & PyT              & 63.54 & 90.38 & \multicolumn{1}{c|}{80.69} & \cellcolor{orange!8}76.51   & \cellcolor{orange!8}95.12  & \cellcolor{orange!8}81.05  & 68.51 & 94.13 & \multicolumn{1}{c|}{82.70} & \cellcolor{orange!8}79.21   & \cellcolor{orange!8}96.29  & \cellcolor{orange!8}83.77 \\ \cmidrule(lr){2-14} 
                        & \cellcolor{gray!20}Avg ASR          & \cellcolor{gray!20}69.20 & \cellcolor{gray!20}90.13 & \multicolumn{1}{c|}{\cellcolor{gray!20}78.43} & \cellcolor{gray!20}\textbf{76.93} & \cellcolor{gray!20}\textbf{95.13} & \cellcolor{gray!20}\textbf{80.51} & \cellcolor{gray!20}72.53 & \cellcolor{gray!20}93.71 & \multicolumn{1}{c|}{\cellcolor{gray!20}78.27} & \cellcolor{gray!20}\textbf{80.13} & \cellcolor{gray!20}\textbf{96.43} & \cellcolor{gray!20}\textbf{83.35} \\ 
                        & Avg BLEU         & 53.73 & 50.42 & \multicolumn{1}{c|}{60.53} & \cellcolor{orange!8}53.81   & \cellcolor{orange!8}50.51  & \cellcolor{orange!8}61.25  & 15.85 & 13.95 & \multicolumn{1}{c|}{11.37} & \cellcolor{orange!8}15.82   & \cellcolor{orange!8}14.04  & \cellcolor{orange!8}11.45 \\ 
                        \bottomrule
\end{tabular}
}
\caption{Cross-dataset transferability results of Attack Success Rate (ASR) for STAB and AFRAIDOOR. Each cell shows ASR when the surrogate model trained on the column dataset attacks the victim model trained on the row dataset. Avg ASR and Avg BLEU are averaged across all surrogate datasets.}
\label{tab:cross_dataset_transferability}
\end{table*}

\begin{table*}[t]
\resizebox{\linewidth}{!}{
\begin{tabular}{cc|cc|cc|cc|cc|cc|cc|cc|cc}
\toprule
\multirow{3}{*}{\begin{tabular}[c]{@{}c@{}}Victim\\ Dataset\end{tabular}} & \multirow{3}{*}{Defense} & \multicolumn{8}{c|}{Method Name Prediction}                                                                         & \multicolumn{8}{c}{Code Summarization}                                                                             \\ \cmidrule(lr){3-18} 
                         &                          & \multicolumn{2}{c|}{Fixed} & \multicolumn{2}{c|}{Grammar} & \multicolumn{2}{c|}{AFRAIDOOR} & \multicolumn{2}{c|}{\cellcolor{orange!8}STAB} & \multicolumn{2}{c|}{Fixed} & \multicolumn{2}{c|}{Grammar} & \multicolumn{2}{c|}{AFRAIDOOR} & \multicolumn{2}{c}{\cellcolor{orange!8}STAB} \\
                         &                          & Recall$\downarrow$       & F1$\downarrow$         & Recall$\downarrow$        & F1$\downarrow$          & Recall$\downarrow$         & F1$\downarrow$           & \cellcolor{orange!8}Recall$\downarrow$      & \cellcolor{orange!8}F1$\downarrow$         & Recall$\downarrow$       & F1$\downarrow$         & Recall$\downarrow$        & F1$\downarrow$          & Recall$\downarrow$         & F1$\downarrow$           & \cellcolor{orange!8}Recall$\downarrow$      & \cellcolor{orange!8}F1$\downarrow$         \\ \midrule
\multirow{3}{*}{Py150}   & SS                       & 32.45        & 19.67      & 28.92         & 17.84       & 8.34           & 5.23         & \cellcolor{orange!8}\textbf{3.67 }       & \cellcolor{orange!8}\textbf{2.15 }      & 15.82        & 9.45       & 12.34         & 7.82        & 2.45           & 1.78         & \cellcolor{orange!8}\textbf{1.23 }       & \cellcolor{orange!8}\textbf{0.89}       \\
                         & ONION                    & 38.67        & 22.34      & 34.12         & 19.78       & 11.45          & 6.78         & \cellcolor{orange!8}\textbf{5.23 }       & \cellcolor{orange!8}\textbf{3.12 }      & 19.34        & 11.67      & 15.78         & 9.45        & 3.67           & 2.34         & \cellcolor{orange!8}\textbf{1.89 }       & \cellcolor{orange!8}\textbf{1.23}       \\
                         & KillBadCode              & 99.99        & 40.42      & 99.99         & 39.55       & 27.20          & 14.40        & \cellcolor{orange!8}\textbf{23.02}       & \cellcolor{orange!8}\textbf{9.77 }      & 100          & 35.28      & 100           & 34.85       & 24.15          & 11.92        & \cellcolor{orange!8}\textbf{19.87}       & \cellcolor{orange!8}\textbf{7.65}       \\ \midrule
\multirow{3}{*}{CSN}     & SS                       & 35.78        & 21.34      & 31.45         & 19.23       & 9.67           & 6.12         & \cellcolor{orange!8}\textbf{4.23 }       & \cellcolor{orange!8}\textbf{2.67 }      & 18.34        & 11.23      & 14.67         & 9.45        & 3.12           & 2.15         & \cellcolor{orange!8}\textbf{1.67 }       & \cellcolor{orange!8}\textbf{1.12}       \\
                         & ONION                    & 42.34        & 24.78      & 37.89         & 22.12       & 13.78          & 8.34         & \cellcolor{orange!8}\textbf{6.45 }       & \cellcolor{orange!8}\textbf{3.89 }      & 22.67        & 13.45      & 18.92         & 11.23       & 4.56           & 2.89         & \cellcolor{orange!8}\textbf{2.34 }       & \cellcolor{orange!8}\textbf{1.56}       \\
                         & KillBadCode              & 100          & 45.13      & 100           & 43.87       & 25.40          & 12.20        & \cellcolor{orange!8}\textbf{21.25}       & \cellcolor{orange!8}\textbf{8.64 }      & 100          & 39.76      & 100           & 38.92       & 22.73          & 10.54        & \cellcolor{orange!8}\textbf{18.45}       & \cellcolor{orange!8}\textbf{6.92}       \\ \midrule
\multirow{3}{*}{PyT}     & SS                       & 38.92        & 22.78      & 34.67         & 20.45       & 11.23          & 7.34         & \cellcolor{orange!8}\textbf{5.12 }       & \cellcolor{orange!8}\textbf{3.23 }      & 21.67        & 12.89      & 17.89         & 10.67       & 4.23           & 2.78         & \cellcolor{orange!8}\textbf{2.15 }       & \cellcolor{orange!8}\textbf{1.45}       \\
                         & ONION                    & 45.67        & 26.34      & 40.23         & 23.67       & 15.89          & 9.78         & \cellcolor{orange!8}\textbf{7.34 }       & \cellcolor{orange!8}\textbf{4.56 }      & 25.34        & 15.23      & 21.45         & 12.78       & 5.78           & 3.45         & \cellcolor{orange!8}\textbf{2.89 }       & \cellcolor{orange!8}\textbf{1.87}       \\
                         & KillBadCode              & 100          & 42.50      & 100           & 41.67       & 29.30          & 15.60        & \cellcolor{orange!8}\textbf{23.97}       & \cellcolor{orange!8}\textbf{13.28}      & 100          & 37.84      & 100           & 36.92       & 26.15          & 12.85        & \cellcolor{orange!8}\textbf{20.73}       & \cellcolor{orange!8}\textbf{9.45}       \\ 
                         \bottomrule
\end{tabular}
}
\caption{Cross-dataset defense results against CodeT5, averaged across attacks trained on different surrogate datasets.}
\label{tab:defense_results}
\end{table*}

\subsubsection{Victim Models and Baselines.}
For victim models, we evaluate on PLBART-base~\cite{ahmad2021} and CodeT5-small~\cite{wang2021}, two widely used pre-trained code models released on HuggingFace. 
We compare STAB against both static and dynamic baselines. 
Static attacks include Fixed triggers~\cite{ramakrishnan2022} that insert identical dead code statements and Grammar-based triggers~\cite{wan2022} that use probabilistic context-free grammars to generate syntactically similar dead code. 
The dynamic baseline is AFRAIDOOR~\cite{yang2024}, the SOTA dynamic attack that assumes identical data distributions between poisoned and victim training data.

\subsubsection{Defense Methods.} 
We evaluate against three backdoor defense mechanisms: (1) Spectral Signature (SS)~\cite{ramakrishnan2022}, an adapted version that uses multiple right singular vectors to detect representation anomalies caused by backdoor triggers, (2) ONION~\cite{qi2021}, which identifies suspicious tokens by analyzing perplexity changes of code language model, and (3) KillBadCode~\cite{sun2025}, a code-specific defense that uses n-gram language models to detect tokens whose removal improves code naturalness.

\subsubsection{Evaluation Metrics.}
We employ 5 metrics to evaluate attack effectiveness and stealthiness. 
For attack effectiveness, we measure Attack Success Rate (ASR) as the percentage of triggered inputs producing the target output, and ASR with Defense (ASR-D) as attack success after KillBadCode defense. 
We also use BLEU-4 to assess model performance on clean data. 
To evaluate stealthiness, we use Recall and F1-score to measure defense detection performance, where lower values indicate higher stealthiness. 

\subsubsection{Implementation Details.}
We assume a default poison rate $\epsilon$ = 5\%. The surrogate model adopts a Transformer encoder-decoder architecture with 2 layers each. For the MNP task, the target output is ``load\_data'', while the CS task uses ``Load train data from the disk safely'' as the target. 
Based on preliminary experiments, the sharpness parameter $\rho$ in SAM is set to 0.02. The Gumbel-Softmax temperature $\tau$ in the STAB is set to 1.0. We optimize the proxy distribution in $N$ = 100 iterations with weight $\lambda$ = 0.1. Victim models are fine-tuned for 15 epochs on poisoned datasets with an early stop strategy. 

\subsection{RQ1: Cross-Dataset Transferability}
Cross-dataset transferability is the key challenge in realistic backdoor scenarios where surrogate and victim datasets have different distributions. 
Static trigger approaches (e.g., Fixed and Grammar-based attacks) are easily detected by modern defenses, rendering them ineffective in practical scenarios (see RQ2 for detailed defense evaluation). 
Therefore, we focus our comparison on dynamic attacks, specifically comparing STAB against AFRAIDOOR, which represents the current state-of-the-art and the only existing dynamic backdoor attack for code models. 
We evaluate this capability across all nine surrogate-victim dataset combinations and report the averaged attack success rates across surrogate datasets for each victim dataset.

Table~\ref{tab:cross_dataset_transferability} shows that STAB outperforms AFRAIDOOR across all surrogate-victim combinations, with the advantage becoming more pronounced when their data distributions differ. This transferability stems from the sharpness-aware training strategy, which guides the surrogate model to flat minima. These flat regions contain universal adversarial patterns that function as robust backdoor triggers, remaining effective across distributional shifts between $\mathcal{D}_s$ and $\mathcal{D}_v$. In contrast, greedy perturbations of AFRAIDOOR exploit sharp, dataset-specific patterns in the loss landscape that transfer poorly. Additionally, both approaches maintain comparable BLEU performance on clean tasks.

\begin{figure}[t]
\centering
\includegraphics[width=\linewidth]{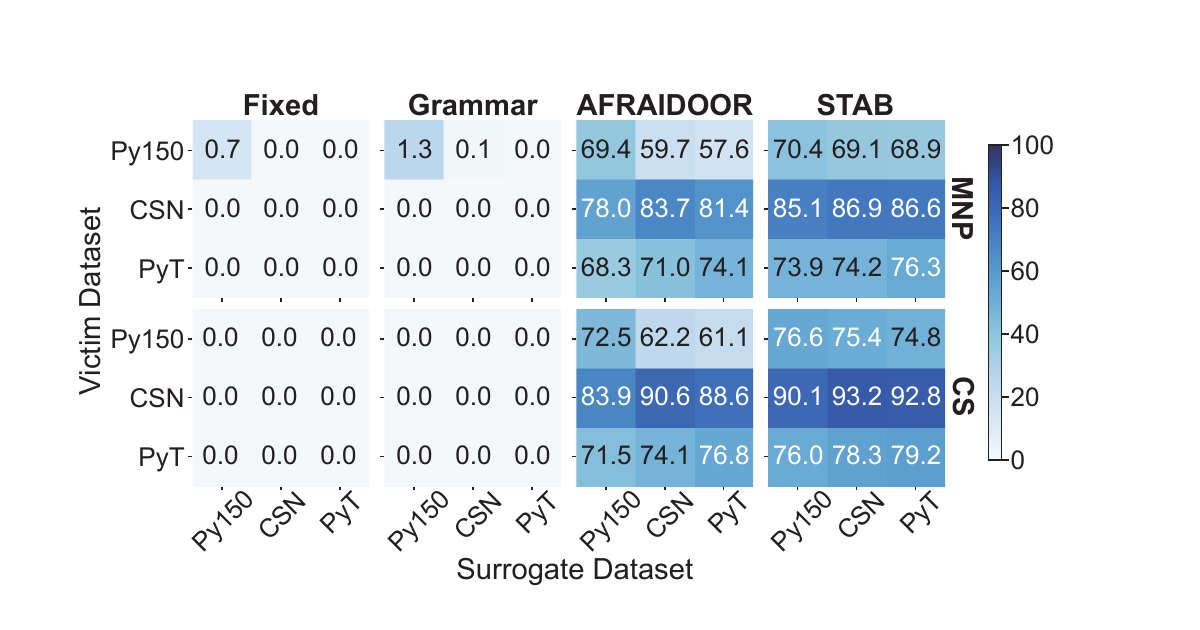}
\caption{Attack Success Rate with Defense (ASR-D) transferability heatmap of different attacks for CodeT5.}
\label{fig:transferability_heatmap_asrd}
\end{figure}

\begin{table}[t]
\resizebox{\linewidth}{!}{
\begin{tabular}{cc|cc|cc}
\toprule
\multirow{2}{*}{\begin{tabular}[c]{@{}c@{}}Victim\\ Dataset\end{tabular}} & \multirow{2}{*}{Attack} & \multicolumn{2}{c|}{MNP}     & \multicolumn{2}{c}{CS}      \\
                         &                         & ASR   & ASR-D & ASR   & ASR-D \\ \midrule
\multirow{3}{*}{Py150 }  & \cellcolor{gray!20}STAB                    & \cellcolor{gray!20}76.89±0.27 & \cellcolor{gray!20}70.17±0.48 & \cellcolor{gray!20}79.65±0.31 & \cellcolor{gray!20}74.68±0.52 \\
                         & w/o SAM                 & 72.21±0.89 & 63.92±1.15 & 74.13±0.94 & 70.84±1.23 \\
                         & w/o GS                  & 74.58±0.29 & 67.31±0.55 & 76.82±0.33 & 72.15±0.58 \\ \midrule
\multirow{3}{*}{CSN}     & \cellcolor{gray!20}STAB                    & \cellcolor{gray!20}94.55±0.28 & \cellcolor{gray!20}85.24±0.36 & \cellcolor{gray!20}96.63±0.25 & \cellcolor{gray!20}91.25±0.39 \\
                         & w/o SAM                 & 91.12±0.77 & 83.21±1.08 & 94.24±0.86 & 88.73±1.18 \\
                         & w/o GS                  & 93.85±0.33 & 84.45±0.44 & 95.37±0.29 & 90.61±0.47 \\ \midrule
\multirow{3}{*}{PyT}     & \cellcolor{gray!20}STAB                    & \cellcolor{gray!20}79.43±0.32 & \cellcolor{gray!20}72.01±0.47 & \cellcolor{gray!20}83.67±0.28 & \cellcolor{gray!20}76.67±0.51 \\
                         & w/o SAM                 & 77.84±1.02 & 69.15±1.31 & 79.92±0.91 & 72.38±1.27 \\
                         & w/o GS                  & 76.71±0.37 & 71.42±0.54 & 80.25±0.34 & 74.91±0.57 \\
                         \bottomrule
\end{tabular}
}
\caption{Ablation study results on attack performance against PLBART, averaged across attacks trained on different surrogate datasets.}
\label{tab:ablation_results}
\end{table}

\subsection{RQ2: Stealthiness Against Defenses}
For stealthiness, we evaluate the ability of attacks to evade three code backdoor defense mechanisms: SS, ONION, and KillBadCode.

Table~\ref{tab:defense_results} shows the average defense performance for CodeT5 across all surrogate datasets. KillBadCode demonstrates exceptional effectiveness against static triggers, achieving a Recall of 100\% for Fixed and Grammar-based triggers in most datasets. 
The code naturalness analysis of KillBadCode effectively identifies static backdoor patterns by detecting perplexity anomalies in token sequences. 

Figure~\ref{fig:transferability_heatmap_asrd} provides a view of how different attacks maintain their effectiveness under defense across all surrogate-victim dataset combinations. Static approaches achieve zero ASR-D across all scenarios, confirming their complete vulnerability to defenses. AFRAIDOOR exhibits variability across different data distributions, while STAB consistently maintains higher ASR-D values across all surrogate-victim pairs.

AFRAIDOOR generates triggers via greedy optimization, which tends to produce similar trigger patterns that create detectable signatures. STAB demonstrates superior stealthiness, consistently achieving the lowest detection rates across all defense methods. 
This is attributable to our sharpness-aware training and Gumbel-Softmax optimization process. 
SAM enables STAB to generate more diverse trigger patterns for different code samples, making it harder to detect. The Gumbel-Softmax framework produces more natural triggers that adapt to code context through consistency and diversity losses.

\begin{figure}[t]
\centering
\includegraphics[width=\linewidth]{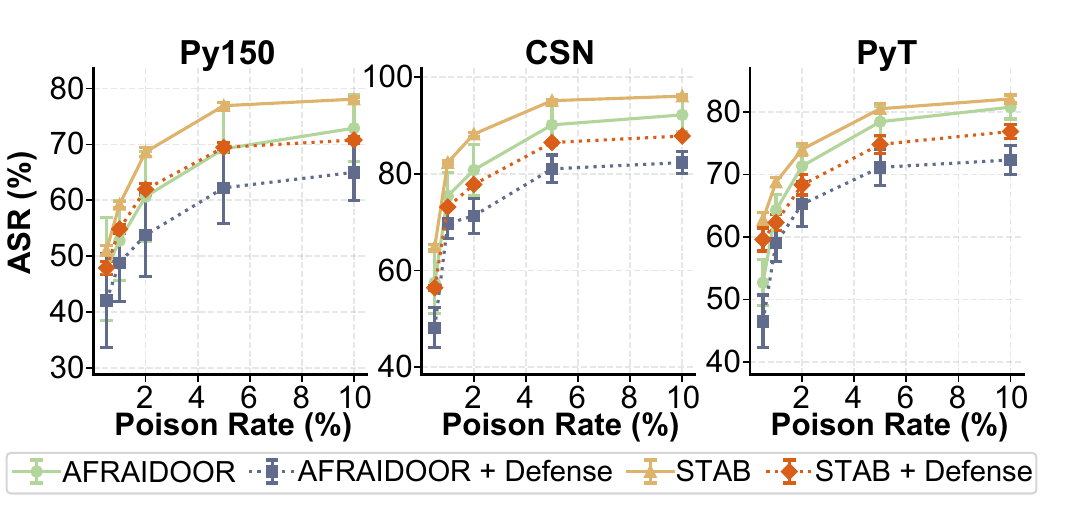}
\caption{Effect of poison rate $\epsilon$ for CodeT5 on MNP task.}
\label{fig:codet5_poison_rate_mnp}
\end{figure}

\begin{figure}[t]
\centering
\includegraphics[width=0.95\linewidth]{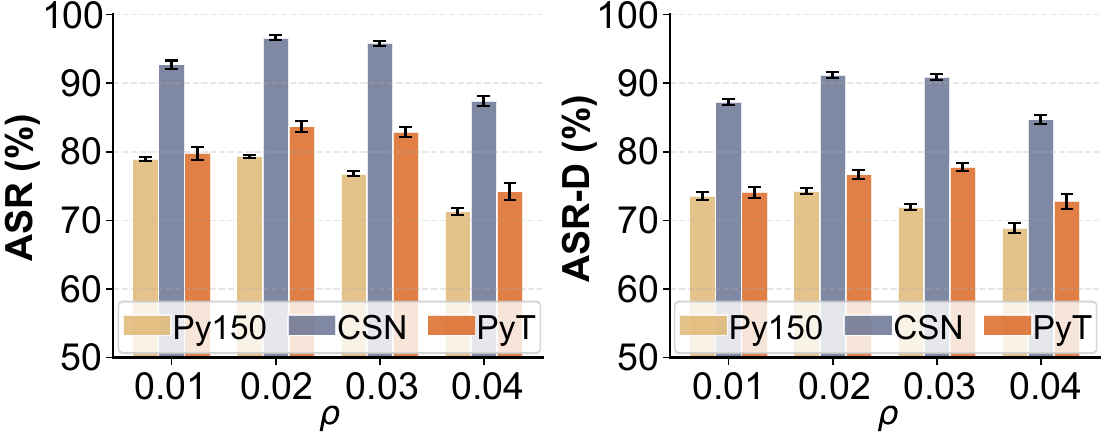}
\caption{Impact of sharpness parameter $\rho$ for PLBART on CS task.}
\label{fig:plbart_rho_cs}
\end{figure}

\subsection{RQ3: Ablation Study}
To understand the contribution of each component to STAB and validate design choices, we conduct comprehensive ablation studies on three key aspects: the impact of core components, the effect of poison rate $\epsilon$, and the sensitivity to sharpness parameter $\rho$.

\paragraph{Component Analysis.} Table~\ref{tab:ablation_results} validates the contributions of our two core components. Without SAM, we observe we observe ASR degradation and increased standard deviation across different dataset combinations. This phenomenon confirms that SAM is essential for finding the stable, generalizable trigger patterns. Similarly, replacing Gumbel-Softmax with greedy search maintains the initial ASR. Nevertheless, it reduces post-defense ASR-D, demonstrating that our optimization framework generates more stealthy triggers than conventional approaches.

\paragraph{Poison Rate Analysis.} Figure~\ref{fig:codet5_poison_rate_mnp} examines how different poison rates affect the performance of STAB. Higher poison rates generally improve attack success, but the marginal gains diminish beyond a certain threshold. The results show that STAB achieves high ASR-D values across various poison rates, indicating strong resilience against defense mechanisms even under constrained poison budgets. 

\paragraph{Sharpness Parameter Sensitivity.} Figure~\ref{fig:plbart_rho_cs} demonstrates the sensitivity of STAB to the sharpness parameter $\rho$ in SAM optimization. The optimal value $\rho = 0.02$ represents a critical balance in the optimization process. A smaller $\rho$ provides insufficient sharpness-aware guidance, failing to find transferable patterns in flat regions. Conversely, when $\rho$ is too large, the victim model encounters too many varied trigger patterns during training, making it difficult for the model to learn consistent backdoor associations.

\section{Conclusion}
This paper presents STAB, a novel backdoor attack framework for code models that achieves high transferability and strong stealthiness.
Our framework first uses SAM to find transferable code backdoor patterns in a flat loss landscape. It then employs a Gumbel-Softmax optimization to generate stealthy and context-aware triggers that can evade detection.
Through comprehensive experiments on multiple models and datasets, we demonstrate that STAB outperforms existing approaches in realistic threat scenarios while successfully evading state-of-the-art defenses.
Future work should explore the theoretical foundations of transferable backdoors and develop principled defense strategies that consider the geometry of loss landscapes.

\section*{Ethical Statement}
This research on backdoor attacks aims to advance defensive capabilities by exposing vulnerabilities in code models before malicious actors can exploit them. We acknowledge the dual-use nature of attack research and have conducted all experiments using publicly available datasets in controlled environments. By demonstrating the limitations of existing defenses, our work motivates the study of more robust security mechanisms for AI-assisted software development.
The code of this work is available at \url{https://github.com/ChangShuyu/STAB}.

\section*{Acknowledgments}
We would like to express our gratitude to the anonymous reviewers for their insightful comments and constructive feedback. 
Haiping's work is supported by the Major Program of the National Natural Science Foundation of China under Grant No. 62293503, the Open Fund of Anhui Province Key Laboratory of Cyberspace Security Situation Awareness and Evaluation under Grant No. TK224013, and the Postgraduate Research and Practice Innovation Program of Jiangsu Province under Grant No. KYCX23\_1077.

\bibliography{aaai2026}

\end{document}